\documentclass[aps,prd,onecolumn,groupedaddress,showpacs,nofootinbib,amssymb
]{revtex4}
\usepackage[dvips]{graphicx}
\usepackage{amssymb}
\usepackage{amsmath}
\usepackage{graphicx}
\usepackage{amsfonts}
\usepackage{bm}

\begin{document}

\title{Inflationary $\alpha$-attractors from $F(R)$ Gravity}
\author{
S.~D.~Odintsov,$^{1,2}$\,\thanks{odintsov@ieec.uab.es}
V.~K.~Oikonomou,$^{3,4}$\,\thanks{v.k.oikonomou1979@gmail.com}}
\affiliation{ $^{1)}$ICREA, Passeig Luis Companys,
23, 08010 Barcelona, Spain\\
$^{2)}$ Institute of Space Sciences (IEEC-CSIC)\\
C. Can Magrans s/n, 08193 Barcelona, SPAIN\\
$^{3)}$ Tomsk State Pedagogical University, 634061 Tomsk, Russia\\
$^{4)}$ Laboratory for Theoretical Cosmology, Tomsk State University
of Control Systems
and Radioelectronics (TUSUR), 634050 Tomsk, Russia\\
}

\begin{abstract}
In this paper we study some classes of $\alpha$-attractors models in
the Jordan frame and we find the corresponding $F(R)$ gravity
theory. We study analytically the problem at leading order and we
investigate whether the attractor picture persists in the $F(R)$
gravity equivalent theory. As we show, if the slow-roll conditions
are assumed in the Jordan frame, the spectral index of primordial
curvature perturbations and the scalar-to-tensor ratio are identical
to the corresponding observational indices of the $R^2$ model, a
result which indicates that the attractor property is also found in
the corresponding $F(R)$ gravity theories of the $\alpha$-attractors
models. Moreover, implicit and approximate forms of the $F(R)$
gravity inflationary attractors are found.
\end{abstract}


\maketitle



\def\pp{{\, \mid \hskip -1.5mm =}}
\def\cL{\mathcal{L}}
\def\be{\begin{equation}}
\def\ee{\end{equation}}
\def\bea{\begin{eqnarray}}
\def\eea{\end{eqnarray}}
\def\tr{\mathrm{tr}\, }
\def\nn{\nonumber \\}
\def\e{\mathrm{e}}

\section{Introduction}

Inflationary cosmology is one of the two existing descriptions of
the early Universe, in the context of which the theoretical
inconsistencies of the Big Bang description of our Universe were
successfully addressed
\cite{inflation1,inflation2,inflation3,inflation4}, with the other
alternative being bouncing cosmology
\cite{bounce1,bounce2,bounce3,bounce4,bounce5,bounce6}. The latest
observational data coming from Planck \cite{planck} posed stringent
constraints on inflationary models and verified the validity of some
models, while rendering other models non-viable.

Recently, an interesting class of models was discovered in
\cite{alpha1}, called the $\alpha$-attractors models, with the
characteristic property of these models being that the predicted
spectral index of primordial curvature perturbations and the
scalar-to-tensor ratio was identical for all the models, in the
large $N$ limit, where $N$ is the $e$-foldings number. These models
were later on studied in
\cite{alpha2,alpha3,alpha4,alpha5,alpha6,alpha7,alpha8,alpha9,alpha10},
and also for an earlier study of some $\alpha$-attractor-like
potentials, see Ref. \cite{slowrollsergei}. Well known inflationary
models are special limiting cases of some $\alpha$ attractors
models, like for example the Starobinsky model
\cite{starob1,starob2}, the Higgs inflationary model \cite{higgs}.
To our opinion the most appealing property of the
$\alpha$-attractors models is that these models have a large plateau
in their potential, for large scalar field values, and in the small
$\alpha$ limit, and the potential of these models is asymptotically
quite similar to the hybrid inflation scenario \cite{hybrid}. The
hybrid inflation model introduced flat potentials in the research
field of inflationary cosmology, and a crucial assumption was used,
that the initial state of the scalar field did not correspond to the
extremum of the scalar potential. Effectively, in the large field
limit, all the $\alpha$-attractors models tend to some variant form
of the hybrid inflation scenario, hence the very own idea of hybrid
inflation is successful, in view of the observational data. The
$\alpha$-attractors models originate from supergravity models, which
also involve a controllable supersymmetry breaking at the minimum of
the scalar potential \cite{susybr1}. In addition, in some cases,
late-time acceleration can be accommodated in the context of
$\alpha$-attractors models \cite{linderefs1,linder}.

In this paper we aim to investigate whether the attractor property
also occurs in the Jordan frame, in terms of the corresponding
$F(R)$ gravity. In principle, since the curvature perturbation in
the Jordan frame is invariant under a conformal transformation, and
also since the tensor perturbations are invariant, someone would
expect that the both the spectral index and also the
scalar-to-tensor ratio in the Jordan frame, should be identical to
the ones calculated for the Einstein frame. However, in the large
field limit, the conformal transformation diverges, so in principle,
the era where the scalar field reaches the plateau, makes the
conformal transformation unbounded. This large field era corresponds
to a pole in the scalar-field Jordan frame, and it is exactly the
era where the $\alpha$-attractors models yield equivalent
observational data. Motivated by this, the focus in this paper is to
check explicitly whether the attractors property is shared by the
$F(R)$ gravity equivalent of these theories. We shall use the large
field limit of the $\alpha$-attractors models and using well-known
methods we shall find the Jordan frame $F(R)$ gravity theory
corresponding to the potential of the $\alpha$-attractors models. As
we explicitly demonstrate, the general problem is not so trivial as
it seems, since the analytic treatment is in general impossible to
be performed, for general values of $\alpha$. In effect, we
investigate certain convenient examples and also for various limits
of the parameter $\alpha$, and as we show, the attractors property
holds true for the $F(R)$ gravity equivalent theories, and more
importantly, the models yield identical observational data to the
$R^2$ inflation model. This finds its explanation to the fact that
the cosmological evolution during the slow-roll era is a quasi-de
Sitter evolution.

But why there is a need to study the physics in different frames?
This is a deep question, so now we shall try to answer this
question, since this is our main motivation for the subject of this
paper. In general, for every theoretical proposal in modified
gravity, it is compelling to compare the results with the
observational data. In this research line, the $F(R)$ gravity Jordan
frame and/or the Einstein frame, may provide a viable description of
the observable Universe, however it is not for sure that a viable
theory in the Jordan frame may give also a viable theory in the
Einstein frame. In addition, the viability of a theoretical
description does not come in hand with the physically convenient
description. So the question is which of the two frames is the more
physical one (at least, in some sense), or  which of the two frames
describes in a more appealing way the cosmic history of our
Universe. To a great extent, the answer to this question depends on
the compatibility of the resulting theory with the observational
data. In addition, in principle there are quantities that should be
the same in the Jordan and Einstein frames, and these are actually
the quantities that are invariant under conformal transformations.
For a quasi-de Sitter evolution, it is expected that the spectral
index and the scalar-to-tensor ratio should be equivalent in the two
frames, as it was shown in \cite{kaizer,newsergei}. However, this
should be explicitly checked, since when neutron stars are studied,
different results occur in the two frames \cite{capp}. In addition,
a finite-time singularity of a certain type in one frame, does not
correspond to the same type of singularity in the other frame
\cite{noo5,bahamonte}, since the conformal transformation becomes
ill defined on the singular point. Also the presence of matter can
lead to escalated complications between frames, since it is
minimally coupled in the Jordan frame but it is non-minimally
coupled in the corresponding Einstein frame. Also it may occur that
the Universe is accelerating in one frame, but decelerates in the
other \cite{capp2}. Hence, these arguments essentially explain our
motivation to study the attractor picture in the Jordan frame.

This paper is organized as follows: In section II we describe in
brief the essential features of the $\alpha$-attractors models, and
we demonstrate how the attractor property occurs. In section III, we
address the same issue in the $F(R)$ gravity equivalent theory and
we demonstrate in detail how the attractor property occurs in this
case, by studying analytically some characteristic examples and
limiting cases. Finally, the conclusions follow in the end of the
paper.

Also in this paper we will assume that the geometric background will
be a flat Friedmann-Robertson-Walker metric, with line element, \be
\label{metricfrw} ds^2 = - dt^2 + a(t)^2 \sum_{i=1,2,3}
\left(dx^i\right)^2\, , \ee with $a(t)$ denoting as usual the scale
factor. Moreover, we assume that the connection is a symmetric,
metric compatible and torsion-less affine connection, the so-called
Levi-Civita connection. For the metric with line element that of Eq.
(\ref{metricfrw}), the Ricci scalar reads,
\begin{equation}
\label{ricciscal} R=6(2H^2+\dot{H})\, ,
\end{equation}
with $H$ denoting the Hubble rate $H=\dot{a}/a$. Also we use a units
system such that $\hbar=c=8\pi G=\kappa^2=1$.

\section{The Inflationary Attractors Essentials and the $F(R)$ Gravity Description}

As we mentioned in the introduction, the terminology
$\alpha$-attractor models refers to inflationary models with plateau
potentials
\cite{alpha1,alpha2,alpha3,alpha4,alpha5,alpha6,alpha7,alpha8,alpha9,slowrollsergei}.
These models include the $R^2$ inflation model in the Einstein frame
\cite{starob1,starob2}, and the Higgs inflation model \cite{higgs}.
An essential feature in these models is the existence of a pole in
the kinetic term of the non-canonical scalar field description.
Usually the description using a non-canonical scalar field is called
the Jordan frame description, so in order to avoid confusion with
the $F(R)$ description, we shall refer to the non-canonical scalar
field Jordan frame as ``$\phi$-Jordan frame'' and to the $F(R)$
Jordan frame simply as ``Jordan frame''.

In the $\phi$-Jordan frame, the $\alpha$-attractors models have the
following gravitational action \cite{alpha1},
\begin{equation}\label{alphaact}
\mathcal{S}=\sqrt{-g}\left(\frac{R}{2}-\frac{\partial_{\mu}\phi
\partial^{\mu}\phi}{2(1-\frac{\phi^2}{6\alpha})^2} -V(\phi)\right)\,
,
\end{equation}
where $R$ is the Ricci scalar and also we used units where the
gravitational constant is $G=1$. Notice that the action
(\ref{alphaact}) contains a pole at $\phi=\sqrt{6\alpha}$, and this
is of fundamental importance in the $\alpha$-attractor theories,
since the order of the pole crucially affects the spectral index of
primordial curvature perturbations $n_s$, while the residue of the
pole affects the scalar-to-tensor ratio $r$ \cite{alpha4}. By making
the transformation,
\begin{equation}\label{dftrans}
\frac{\mathrm{d}\phi }{1-\frac{\phi^2}{6\alpha}}=\mathrm{d}\varphi\,
,
\end{equation}
the non-canonical action of Eq. (\ref{alphaact}) is transformed into
the canonical scalar field action,
\begin{equation}\label{canonocclact}
\mathcal{S}=\sqrt{-g}\left(\frac{R}{2}-\frac{1}{2}\partial_{\mu}\phi
\partial^{\mu}\phi -V(\sqrt{6\alpha}\tanh
(\frac{\varphi}{\sqrt{6\alpha}}))\right)\, ,
\end{equation}
where the argument of the scalar potential easily follows by solving
the transformation equation (\ref{dftrans}).

One of the most interesting features of the $\alpha$-attractors
models is that at small $\alpha$, or equivalently at large $\varphi$
values, the quite generic potentials $V(\sqrt{6\alpha}\tanh
(\frac{\varphi}{\sqrt{6\alpha}}))$ approach an infinitely long de
Sitter plateau, which corresponds to the value of the non-canonical
potential $V(\phi)$ at the boundary $V(\phi )\Big{|}_{\pm
\sqrt{6\alpha}}$. The terminology attractors is justified due to the
fact that regardless of the form of the potential, all the
$\alpha$-attractor models lead to the same spectral index of
primordial curvature perturbations $n_s$ and to the same
scalar-to-tensor ratio $r$, in the small $\alpha$ limit, which have
the following form,
\begin{equation}\label{scaspectscalar}
n_s\simeq 1-\frac{2}{N},\,\,\,r\simeq \frac{12\alpha}{N^2}\, ,
\end{equation}
where $N$ is the $e$-foldings number. The purpose of this paper is
to investigate if this attractor picture remains when one considers
the $F(R)$ gravity equivalent theory corresponding to the canonical
scalar fields. The main reason behind the attractor picture in the
Einstein frame is that the various generic potentials
$V(\sqrt{6\alpha}\tanh (\frac{\varphi}{\sqrt{6\alpha}}))$ have a
similar limiting behavior in the small $\alpha$ limit. We shall
consider two classes of potentials, namely the T-models and the
E-models, with the potential in the T-models case being of the
following form,
\begin{equation}\label{tmodels}
V(\varphi)=\alpha \mu^2 \tanh^2 (\frac{\varphi}{\sqrt{6\alpha}})\, ,
\end{equation}
where $\mu$ is a positive number, freely chosen. In the case of the
E-models, the potential has the following form,
\begin{equation}\label{potentialemodels}
V(\varphi )=\alpha \mu^2 \left(
1-e^{-\sqrt{\frac{2}{3\alpha}}\varphi}\right)^{2 n}\, ,
\end{equation}
with the parameter $n$ being a positive number, not necessarily an
integer. Note that for $\alpha=n=1$, the potential
(\ref{potentialemodels}) becomes,
\begin{equation}\label{starobmodel}
V(\varphi )=\alpha \mu^2 \left(
1-e^{-\sqrt{\frac{2}{3}}\varphi}\right)^{2}\, ,
\end{equation}
which is the Starobinsky model \cite{starob1}, so essentially the
Starobinsky model is a subcase of the E-models. The scalar potential
in the large $\varphi$ limit becomes approximately equal to,
\begin{equation}\label{limtmodel}
V(\varphi )\simeq \alpha \mu^2 \left(1-
4e^{-\sqrt{\frac{2}{3\alpha}}\varphi}\right)\, ,
\end{equation}
while the E-model potential in the large $\varphi$ limit becomes
approximately equal to,
\begin{equation}\label{smallalphaemodelpot}
V(\varphi )\simeq \alpha \mu^2 \left(1- 2 n
e^{-\sqrt{\frac{2}{3\alpha}}\varphi}\right)\, .
\end{equation}
As it can be seen, the potentials of Eqs. (\ref{limtmodel}) and
(\ref{smallalphaemodelpot}) coincide when $n=2$, but as we already
mentioned the resulting spectral index $n_s$ and the
scalar-to-tensor ratio coincide for general $n$, and also the number
$n$ does not appear in the resulting expressions of $n_s$ and $r$.
Let us briefly demonstrate this issue, since it is of crucial
importance when we compare the Einstein frame observational indices
with the Jordan frame ones. As we will show, any difference should
originate from the slow-roll conditions in the two frames. Let us
consider the limiting case potential of Eq.
(\ref{smallalphaemodelpot}), and in the following we shall focus on
this potential, since almost all the cases we will study result to
this potential in the small $\alpha$ limit. The fist two slow-roll
indices $\epsilon$ and $\eta$ in the slow-roll approximation for a
canonical scalar field are defined as follows,
\begin{equation}\label{slowrollscalar}
\epsilon (\varphi)=\frac{1}{2}\left(
\frac{V'(\varphi)}{V(\varphi)}\right)^2,\,\,\,\eta
(\varphi)=\frac{V''(\varphi)}{V(\varphi)}\, ,
\end{equation}
and also the $e$-foldings number $N$ can also be expressed in terms
of the potential when the slow-roll approximation is used, and it
explicitly reads,
\begin{equation}\label{efoldings}
N\simeq
\int_{\varphi}^{\varphi_{i}}\frac{V(\varphi)}{V'(\varphi)}\mathrm{d}\varphi
\, ,
\end{equation}
where $\varphi_i$ is some initial value of the canonical scalar
field. For the potential (\ref{smallalphaemodelpot}), the
$e$-foldings number $N$ can be expressed in terms of the canonical
scalar field, and in the small $\alpha$ limit, the resulting
expression is,
\begin{equation}\label{slowrollind}
N\simeq \frac{3\alpha e^{\sqrt{\frac{2}{3\alpha }\varphi}}}{4n}\, ,
\end{equation}
so by calculating the slow-roll indices and substituting the
$e$-foldings number from Eq. (\ref{efoldings}), the slow-roll
indices take the following form,
\begin{equation}\label{slowrollapprox1new}
\epsilon\simeq \frac{3\alpha}{4 N^2},\,\,\,\eta \simeq
-\frac{1}{N}\, .
\end{equation}
The spectral index of primordial curvature perturbations $n_s$ and
the scalar-to-tensor ratio $r$ calculated for a canonical scalar
field, are equal to,
\begin{equation}\label{spectscalindex}
n_s\simeq 1-6\epsilon+2\eta,\,\,\, r\simeq 16 \epsilon\, ,
\end{equation}
so by substituting the slow-roll indices from the expressions
(\ref{slowrollapprox1new}), the resulting observational indices are,
\begin{equation}\label{observslowroli1}
n_s\simeq 1-\frac{2}{N}-\frac{9\alpha}{2N^2},\,\,\,r\simeq
\frac{12\alpha}{N^2}\, .
\end{equation}
At large $N$, the observational indices of Eq.
(\ref{observslowroli1}) coincide with the results of Eq.
(\ref{scaspectscalar}), so at leading order only the spectral index
is independent of $\alpha$ and also both the observational indices
do not depend on the parameter $n$. This is exactly the attractor
picture for the general class of the potentials, which have limiting
form (\ref{smallalphaemodelpot}). Below we quote the three crucial
conditions that need to hold true in order the attractor picture in
the Einstein frame occurs:
\begin{itemize}
    \item The small $\alpha$ limit of the potential should be taken.
    \item The large $N$ limit should be taken.
    \item The slow-roll approximation should hold true.
\end{itemize}
As we will show shortly, when these conditions hold true in the
Jordan frame, then the attractor picture occurs in the Jordan frame
too, with the difference that the observational indices have no
$\alpha$ dependence.

Let us start our Jordan frame considerations by firstly finding the
vacuum $F(R)$ gravity \cite{reviews1,reviews2,reviews3,reviews4}
that can generate potentials as in Eq. (\ref{smallalphaemodelpot}).
This limiting case covers both the E-models and T-models in the
small $\alpha$ limit. Before proceeding to the main focus of this
article, we recall some essential features of the connection between
the Einstein and Jordan frame equivalent theories
\cite{reviews1,reviews2,slowrollsergei,sergeioikonomou1,sergeioikonomou2,sergeioikonomou4}.
Consider the following $F(R)$ gravity action,
\begin{equation}\label{pure}
\mathcal{S}=\frac{1}{2}\int\mathrm{d}^4x \sqrt{-\hat{g}}F(R)\, ,
\end{equation}
where $\hat{g}_{\mu \nu}$ is the metric tensor in the Jordan frame.
Introducing the auxiliary field $A$ in the Jordan frame action
(\ref{pure}), the latter can be written as follows,
\begin{equation}\label{action1dse111}
\mathcal{S}=\frac{1}{2}\int \mathrm{d}^4x\sqrt{-\hat{g}}\left (
F'(A)(R-A)+F(A) \right )\, .
\end{equation}
By varying the action of Eq. (\ref{action1dse111}) with respect to
the scalar field $A$, it yields the solution $A=R$, and this proves
the mathematical equivalence of the actions (\ref{action1dse111})
and (\ref{pure}).

A crucial step in finding the Einstein frame canonical scalar-tensor
theory corresponding to the $F(R)$ gravity (\ref{pure}) is to
perform a canonical transformation. It is important to note that the
canonical transformation should not contain the parameter $\alpha$,
see the discussion in the Appendix on this issue. The canonical
transformation that connects the Einstein and Jordan frames is the
following,
\begin{equation}\label{can}
\varphi =\sqrt{\frac{3}{2}}\ln (F'(A))
\end{equation}
with $\varphi$  being the canonical scalar field in the Einstein
frame. Upon conformally transforming the Jordan frame metric
$\hat{g}_{\mu \nu }$ as follows,
\begin{equation}\label{conftransmetr}
g_{\mu \nu}=e^{-\varphi }\hat{g}_{\mu \nu }
\end{equation}
where $g_{\mu \nu}$ denotes the Einstein frame metric, we obtain the
following Einstein frame canonical scalar field action,
\begin{align}\label{einsteinframeaction}
& \mathcal{\tilde{S}}=\int \mathrm{d}^4x\sqrt{-g}\left (
R-\frac{1}{2}\left (\frac{F''(A)}{F'(A)}\right )^2g^{\mu \nu
}\partial_{\mu }A\partial_{\nu }A -\left (
\frac{A}{F'(A)}-\frac{F(A)}{F'(A)^2}\right ) \right ) \\ \notag & =
\int \mathrm{d}^4x\sqrt{-g}\left ( R-\frac{1}{2}g^{\mu \nu
}\partial_{\mu }\varphi\partial_{\nu }\varphi -V(\varphi )\right )
\end{align}
The Einstein frame potential $V(\varphi )$ of the canonical scalar
field $\varphi $ is equal to,
\begin{align}\label{potentialvsigma}
V(\varphi
)=\frac{1}{2}\left(\frac{A}{F'(A)}-\frac{F(A)}{F'(A)^2}\right)=\frac{1}{2}\left
( e^{-\sqrt{2/3}\varphi }R\left (e^{\sqrt{2/3}\varphi} \right )-
e^{-2\sqrt{2/3}\varphi }F\left [ R\left (e^{\sqrt{2/3}\varphi}
\right ) \right ]\right )\, .
\end{align}
The Ricci scalar as a function of the scalar field can be found by
solving Eq. (\ref{can}) with respect to $A$, having in mind of
course the equivalence of $R$ and $A$. It is straightforward to
obtain the $F(R)$ gravity that generates a specific potential, by
simply combining Eqs. (\ref{potentialvsigma}) and (\ref{can}).
Indeed, by taking the derivative of both sides of Eq.
(\ref{potentialvsigma}), with respect to the Ricci scalar, and also
due to the fact that
$\frac{\mathrm{d}\varphi}{\mathrm{d}R}=\sqrt{\frac{3}{2}}\frac{F''(R)}{F'(R)}$,
we obtain the following relation, which is crucial for the analysis
that follows,
\begin{equation}\label{solvequation}
RF_R=2\sqrt{\frac{3}{2}}\frac{\mathrm{d}}{\mathrm{d}\varphi}\left(\frac{V(\varphi)}{e^{-2\left(\sqrt{2/3}\right)\varphi}}\right)
\end{equation}
with $F_R=\frac{\mathrm{d}F(R)}{\mathrm{d}R}$. The above
differential equation (\ref{solvequation}) combined with the
solution of Eq. (\ref{can}) with respect to $R$, will provide us
with the $F(R)$ gravity that generates some of the
$\alpha$-attractors potential we presented earlier. For illustrative
purposes let us see how the $F(R)$ reconstruction method works,
given the Einstein frame. Consider the Starobinsky potential
(\ref{starobmodel}), so by substituting this in Eq.
(\ref{solvequation}), and also using the fact that
$F_R=e^{\sqrt{\frac{2}{3}}\varphi}$, we obtain the following
algebraic equation,
\begin{equation}\label{staroalgebreqn}
F_R R-\left(4 F_R^2 \mu ^2-4 F_R \mu ^2\right)=0\, ,
\end{equation}
which has the solution,
\begin{equation}\label{starsol1}
F_R=\frac{4 \mu ^2+R}{4 \mu ^2}\, ,
\end{equation}
so by integrating with respect to $R$ we obtain the well-known $R^2$
model, which is $F(R)=R+\frac{R^2}{8 \mu^2}$. Note that the latter
result gives implicitly the corresponding $F(R)$-gravity
alpha-attractor.

\section{$F(R)$ Gravity Description: Some Examples for Specific and Limiting Cases of $\alpha$}

\subsection{The Case $\alpha=1/4$}

By using the reconstruction method we presented we will investigate
which $F(R)$ gravities can generate the $\alpha$-attractors
potential we discussed earlier. We shall be interested in the large
$\varphi$ values which correspond to the inflationary de Sitter
plateau in the Einstein frame, or near the pole at $\phi=\sqrt{6}$
in the $\phi$-Jordan frame. Suppose that $\alpha$ is not specified,
so by substituting the potential (\ref{smallalphaemodelpot}) in Eq.
(\ref{solvequation}), we obtain the following algebraic equation,
\begin{equation}\label{algegeneralalpha}
F_R R-4 \alpha  \mu ^2 F_R^{-\left(\sqrt{\frac{1}{\alpha
}}-2\right)} \left(F_R^{\sqrt{\frac{1}{\alpha
}}}+\left(\sqrt{\frac{1}{\alpha }}-2\right) n\right)=0\, .
\end{equation}
However for general $\alpha$ it is a rather formidable task to solve
the algebraic equation (\ref{algegeneralalpha}), so we shall specify
the value of $\alpha$ for various interesting cases. An interesting
case, and one of the few that can be analytically solved, is for
$\alpha=1/4$ since the parameter $\alpha$ is smaller than unity.
Consider that $\alpha=1/4$, in which case the algebraic equation
(\ref{algegeneralalpha}) is simplified as follows,
\begin{equation}\label{casealpha1}
F_R R-F_R^2 \mu ^2=0\, ,
\end{equation}
and the non-trivial solution to (\ref{casealpha1}) is
$F_R(R)=\frac{R}{\mu^2}$, therefore, the resulting $F(R)$ gravity
is,
\begin{equation}\label{casealpjha1solutin1}
F(R)=\frac{R^2}{2\mu^2}+\Lambda\, .
\end{equation}
The integration constant $\Lambda$ can only be specified if we
follow the inverse reconstruction procedure and we identify the
resulting potential with (\ref{smallalphaemodelpot}), for
$\alpha=1/4$. Indeed, by using Eq. (\ref{can}), we obtain that
$R=\mu^2e^{\sqrt{\frac{2}{3}}\varphi}$, so by combining this with
the resulting $F(R)$ gravity (\ref{casealpjha1solutin1}) and by
substituting in the first equation in Eq. (\ref{potentialvsigma}),
we obtain the following potential,
\begin{equation}\label{respot1}
V(\varphi)=\frac{\mu^2}{4}\left(
1-\frac{2\Lambda}{\mu^2}e^{-2\sqrt{\frac{2}{3}}}\right)\, .
\end{equation}
The potential (\ref{respot1}) has to be identical to the one in Eq.
(\ref{smallalphaemodelpot}), so the parameter $\Lambda$ is
$\Lambda=n \mu^2$. Having the $F(R)$ gravity equivalent theory of
the $\alpha$-attractor potential (\ref{smallalphaemodelpot}), we can
calculate the slow-roll indices and the corresponding observational
indices in the Jordan frame and see whether the attractor picture
remains, as in the Einstein frame.

Let us start by finding an approximate expression for the Hubble
rate at early times, as a function of the cosmic time $t$. In order
to do this we will need the cosmological equations for the FRW
metric (\ref{metricfrw}) in the case a general $F(R)$ gravity is
used. By varying the action (\ref{pure}), with respect to the
corresponding metric, we obtain the following cosmological
equations,
\begin{align}\label{cosmoeqns}
& 6F_RH^2=F_RR-F-6H\dot{F}_R ,\\ \notag &
-2\dot{H}F_R=\ddot{F}_R-H\dot{F}_R\, ,
\end{align}
so by using these and the slow-roll approximation, we will be able
to obtain an approximate form for the Hubble rate during the
slow-rolling phase of inflation. We shall use the first equation in
Eq. (\ref{cosmoeqns}), so by substituting the expressions for $F_R$
and $F(R)$ from Eq. (\ref{casealpjha1solutin1}) and also the
expression for the Ricci scalar (\ref{ricciscal}), we obtain the
following differential equation,
\begin{equation}\label{tenfrl}
36 H''(t)+\frac{\mu ^4 n-18 H'(t)^2}{H(t)}+108 H(t) H'(t)=0\, .
\end{equation}
The only dominant term during the slow-roll phase is the last one,
so by solving it we obtain $H(t)=H_0$, which describes a de Sitter
solution. However, it can be checked that the exact de Sitter
solution is not a solution to the following equation,
\begin{equation}\label{dffd}
2F(R)-RF'(R)=0\, ,
\end{equation}
when the $F(R)$ gravity is equal to the one appearing in Eq.
(\ref{casealpjha1solutin1}), so this means that the approximate
solution $H(t)\simeq H_0$ is a leading order result and more terms
are needed in order to better describe the solution. The solution
$H(t)=H_0$ during the slow-roll era where $F'(R)\gg 1$ can also be
verified by using well-known results in the literature
\cite{reviews1}, where for an $F(R)$ gravity of the form
$F(R)=C+\alpha R^n$, $n>0$, the first slow-roll index during the
slow-rolling phase is calculated to be,
\begin{equation}\label{snresult}
\epsilon_1=\frac{2-n}{(n-1)(2n-1)}\, ,
\end{equation}
which is identical to the slow-roll index corresponding to an
$F(R)=R+\alpha R^n$ gravity. Therefore for $n=2$ the first slow-roll
index is zero which implies that $\dot{H}(t)=0$ and hence the Hubble
rate $H(t)\simeq H_0$ describes the evolution during the slow-roll
phase. However, as in the case we described earlier, the exact de
Sitter solution is not a solution to the equation (\ref{dffd}) for
both the $C+\alpha R^n$ and the $R+\alpha R^n$ model, even for
$n=2$. Therefore we seek a leading order quasi-de Sitter evolution,
exactly as in the case of the $R^2$ gravity model. So we
differentiate the first equation in Eq. (\ref{cosmoeqns}), with
respect to the cosmic time, and we get the following differential
equation,
\begin{align}\label{difeqn}
& 6 H'(t) R'(t) F''(R(t))+6 H(t)^2 R'(t) F''(R(t))-R(t) R'(t)
F''(R(t))+6 H(t) \left(R'(t)^2 F'''(R(t))+R''(t) F''(R(t))\right)\\
\notag &+12 H(t) F'(R(t)) H'(t)+F'(R(t)) R'(t)-F'(R(t)) R'(t)=0\, .
\end{align}
In effect, by substituting the $F(R)$ gravity and its higher
derivatives with respect to the Ricci scalar, we obtain the
following differential equation,
\begin{equation}\label{approx2}
\frac{36 H'''(t)}{H(t)}+108 H''(t)+\frac{216 H'(t)^2}{H(t)}=0\,.
\end{equation}
During the slow-roll phase, the first and last terms are
subdominant, since $H(t)\gg H^{(3)}(t) $ and $H'(t)\ll H(t)$, plus
the last term contains a higher power of $H'(t)$. So the only term
that yields the leading order solution is the second term, so by
solving the resulting differential equation we obtain the Hubble
rate during the slow-roll phase which is,
\begin{equation}\label{quasidesitter}
H(t)=H_0-H_i (t-t_k)\, ,
\end{equation}
which is a quasi de Sitter evolution, and $H_0$, $H_i$ are arbitrary
integration constants. Note that $t_k$ is chosen to be the time that
the horizon crossing occurred, at which time the comoving wavenumber
satisfied $k=a(t)H(t)$, with $a(t)$ being the scale factor. Also the
minus sign in the Hubble evolution (\ref{quasidesitter}) has be
chosen in order the first slow-roll parameter at the end of
inflation has a positive sign. Having the approximate expression for
the Hubble rate during the slow-rolling phase, will enable us to
calculate the observational indices in the $F(R)$ gravity case. The
general expressions of the slow-roll indices for an $F(R,\phi)$
gravity with gravitational action (setting $\kappa=1$),
\begin{equation}\label{asx1}
\mathcal{S}=\int \mathrm{d}^4x\sqrt{-g}\left (F(R,\phi)-\frac{\omega
(\phi)}{2}g^{\mu \nu}\partial_{\mu}\phi\partial{\nu}\phi-V(\phi)
\right)\, ,
\end{equation}
are equal to \cite{noh},
\begin{equation}\label{asx2}
\epsilon_1=-\frac{\dot{H}}{H^2},\,\,\,\epsilon_2=\frac{\ddot{\phi}}{H\dot{\phi}},\,\,\,\epsilon_3=\frac{\dot{F}'(R,\phi)}{2
HF'(R,\phi)},\,\,\,\epsilon_4\simeq \frac{\dot{E}}{2HE}\, ,
\end{equation}
where $E$ is equal to,
\begin{equation}\label{epsilonspecif}
E=F'(R,\phi)\omega(\phi)+\frac{3\dot{F}'(R,\phi)^2}{2\dot{\phi}^2}\,
.
\end{equation}
Specifying now, the slow-roll indices as functions of the Hubble
rate for a general $F(R)$ gravity, are equal to \cite{noh},
\begin{equation}\label{slowrollparameters}
\epsilon_1=-\frac{\dot{H}}{H^2},\,\,\,\epsilon_2=0,\,\,\,\epsilon_3=\simeq
\epsilon_1,\,\,\,\epsilon_4\simeq
-3\epsilon_1+\frac{\dot{\epsilon}_1}{H(t)\epsilon_1}\, ,
\end{equation}
and the the spectral index of primordial curvature perturbations
$n_s$ and the scalar-to-tensor ratio $r$ reads,
\begin{equation}\label{nsdf}
n_s\simeq 1-6\epsilon_1-2\epsilon_4\simeq
1-\frac{2\dot{\epsilon}_1}{H(t)\epsilon_1},\,\,\,r=48\epsilon_1^2\,
.
\end{equation}
The slow-roll approximation breaks down at first order when the
first slow-roll parameter becomes of order one, that is
$\epsilon_1\simeq \mathcal{O}(1)$, so at that point, say at $t=t_f$,
assume that the Hubble rate is $H(t_f)=H_f$. From the expression of
the first slow-roll parameter $\epsilon_1$, we get $1\simeq
\frac{H_i}{H_f^2}$, so $H_f\simeq \sqrt{H_i}$. Then from Eq.
(\ref{quasidesitter}) we obtain that,
\begin{equation}\label{finalghf}
H_f-H_0\simeq -H_i (t_f-t_k)\, ,
\end{equation}
so by substituting $H_f$ we get,
\begin{equation}\label{timerelation}
t_f-t_k=\frac{H_0}{H_i}-\frac{\sqrt{H_i}}{H_i}\, .
\end{equation}
Since $H_0$ and $H_i$ are expected to be of the same order during
the slow-roll and also it is expected that these parameters have
quite large values. In effect, the second term in Eq.
(\ref{timerelation}) is subdominant, and therefore we have,
\begin{equation}\label{timerelation1}
t_f-t_k\simeq \frac{H_0}{H_i}\, .
\end{equation}
In order to introduce the $e$-foldings number into the calculation,
we use the relation that expresses the $e$-folding number as a
function of the Hubble rate,
\begin{equation}\label{efold1}
N=\int_{t_k}^{t_f}H(t)\mathrm{d}t\, ,
\end{equation}
calculated from the horizon crossing time until the end of inflation
time. Substituting Eq. (\ref{quasidesitter}) in Eq. (\ref{efold1})
we get,
\begin{equation}\label{nefold1}
N=H_0(t_f-t_k)-\frac{H_i(t_f-t_k)^2}{2}\, ,
\end{equation}
so by using (\ref{timerelation1}) we finally obtain,
\begin{equation}\label{nfinal1}
N=\frac{H_0^2}{2H_i}\, .
\end{equation}
In effect, we have at leading order,
\begin{equation}\label{leadingtf}
t_f-t_k\simeq \frac{2N}{H_0}\, ,
\end{equation}
so by calculating the slow-roll indices, we easily find that the
spectral index and the scalar-to-tensor ratio are equal to,
\begin{equation}\label{spectrscatotensor}
n_s\simeq 1-\frac{4 H_i}{\left(H_0-\frac{2 H_i N}{H_0}\right)^2},\,
\,\,r=\frac{48 H_i^2}{\left(H_0-\frac{2 H_i N}{H_0}\right)^4}\, .
\end{equation}
In the large $N$ limit the observational indices read,
\begin{equation}\label{spectrscatotensor1}
n_s\simeq 1-\frac{H_0^2}{H_i N^2},\, \,\,r=\frac{3 H_0^4}{H_i^2
N^4}\, ,
\end{equation}
so by substituting Eq. (\ref{nfinal1}) in Eq.
(\ref{spectrscatotensor1}) we get,
\begin{equation}\label{jordanframeattract}
n_s\simeq 1-\frac{2}{N},\,\,\,r\simeq \frac{12}{N^2}\, .
\end{equation}
This result is identical to the one of $R^2$-inflation
\cite{starob1} or Higgs inflation \cite{higgs} due to the
well-established equivalence of spectral index and of the
scalar-to-tensor ratio in the Einstein and $F(R)$ frames
\cite{newsergei}, when the slow-roll approximation is assumed.
Therefore, we demonstrated that even for a general value of the
parameter ``$\alpha$'', in the Jordan frame, the general
$\alpha$-$F(R)$ gravity models result to the same spectral index and
scalar-to-tensor ratio, therefore the attractor picture remains in
the Jordan frame too, at least at leading order. We need to note
that a crucial assumption for the calculation was that a slow-roll
era is realized in the model and also the large $N$ limit was taken
in the end. Finally, also note that graceful exit in this theory is
achieved in the same way as in $R^2$ inflation, or may be generated
by growing curvature perturbations.

The calculations we performed here could be performed because the
$\alpha=1/4$ case was easy to deal analytically. However for general
values of $\alpha$ this is not possible. Take for example the
$\alpha=1/9$ case, in which case, the algebraic equation
(\ref{algegeneralalpha}) becomes,
\begin{equation}\label{algbreeqn19}
R F_R-\frac{4 \mu ^2 \left(F_R^3+n\right)}{9 F_R}=0\, ,
\end{equation}
and the solution to this equation is,
\begin{align}\label{sol1}
& F_R= \frac{3 R}{4 \mu ^2}-\frac{9 R^2}{4 \mu ^2 \sqrt[3]{8
\sqrt{16 \mu ^{12} n^2-27 \mu ^6 n R^3}+32 \mu ^6 n-27 R^3}}
-\frac{\sqrt[3]{8 \sqrt{16 \mu ^{12} n^2-27 \mu ^6 n R^3}+32 \mu ^6
n-27 R^3}}{4 \mu ^2}\,.
\end{align}
As it is obvious, this equation cannot be solved analytically with
respect to $R$ and hence, we cannot find an analytic expression for
the potential.

Before closing we need to examine whether the value for $\alpha$ we
chose, namely $\alpha=1/4$, yields viable results in the Einstein
frame. So we examine the Einstein frame observational indices of Eq.
(\ref{observslowroli1}), for $\alpha=1/4$. For the set of values
$(N,\alpha)=(60,1/4)$, we obtain $n_s\simeq 0.966667$ and also
$r\simeq 0.000833333$, which are compatible with the Planck data
which constraint $n_s$ and $r$ as follows \cite{planck},
\begin{equation}
\label{planckdata} n_s=0.9644\pm 0.0049\, , \quad r<0.10\, .
\end{equation}
Also for the set $(N,\alpha)=(50,1/4)$ we get, $n_s\simeq 0.966667$
and also $r\simeq 0.0012$, which are also in agreement with the
Planck data of Eq. (\ref{planckdata}). Hence, for all the physically
relevant values of the $e$-foldings number $N$, which lie in the
interval $N=(50,60)$, the value $\alpha=1/4$ makes the Einstein
frame observables compatible with the Planck data.

\subsection{The Case $\alpha=4$}

Another case that can be treated analytically is the case
$\alpha=4$, in which case the algebraic equation
(\ref{algegeneralalpha})
 becomes,
\begin{equation}\label{casealpha1newsec}
F_R R-16 F_R^{3/2} \mu ^2 \left(\sqrt{F_R}-\frac{3 n}{2}\right)=0\,
,
\end{equation}
and there are two non-trivial solutions to Eq.
(\ref{casealpha1newsec}), which are,
\begin{equation}\label{newfrnontrivs1}
F_R=\frac{18 \mu ^4 n^2+6 \sqrt{9 \mu ^8 n^4+\mu ^6 n^2 R}+\mu ^2
R}{16 \mu ^4}\, ,
\end{equation}
\begin{equation}\label{newfrnontrivs2}
F_R=\frac{18 \mu ^4 n^2-6 \sqrt{9 \mu ^8 n^4+\mu ^6 n^2 R}+\mu ^2
R}{16 \mu ^4}\, ,
\end{equation}
but the only solution which can yield the potential
(\ref{smallalphaemodelpot}) is that of Eq. (\ref{newfrnontrivs1}),
as we now show. Indeed, the $F(R)$ gravity corresponding to Eq.
(\ref{newfrnontrivs1}) is equal to,
\begin{equation}\label{caseonefrnontriv}
F(R)=\frac{9 n^2 \sqrt{\mu ^6 n^2 \left(9 \mu ^2 n^2+R\right)}}{4
\mu ^2}+\frac{R \sqrt{\mu ^6 n^2 \left(9 \mu ^2 n^2+R\right)}}{4 \mu
^4}+\frac{9 n^2 R}{8}+\frac{R^2}{32 \mu ^2}+\Lambda\, ,
\end{equation}
where $\Lambda$ is a constant the value of which will be determined
shortly. Correspondingly, the $F(R)$ gravity corresponding to Eq.
(\ref{newfrnontrivs2}) is equal to,
\begin{equation}\label{caseonefrnontriv1}
F(R)=-\frac{9 n^2 \sqrt{\mu ^6 n^2 \left(9 \mu ^2 n^2+R\right)}}{4
\mu ^2}-\frac{R \sqrt{\mu ^6 n^2 \left(9 \mu ^2 n^2+R\right)}}{4 \mu
^4}+\frac{9 n^2 R}{8}+\frac{R^2}{32 \mu ^2}+\Lambda\, .
\end{equation}
The Einstein frame potential corresponding to the $F(R)$ gravities
above can be easily calculated by using the canonical transformation
relation (\ref{can}), which for both the $F(R)$ gravities of Eqs.
(\ref{caseonefrnontriv}) and (\ref{caseonefrnontriv1}) yields the
following two solutions,
\begin{equation}\label{r1}
R=8 \mu ^2 e^{\frac{\varphi }{\sqrt{6}}} \left(3 n+2
e^{\frac{\varphi }{\sqrt{6}}}\right),
\end{equation}
\begin{equation}\label{r2}
R=8 \mu ^2 e^{\frac{\varphi }{\sqrt{6}}} \left(3 n-2
e^{\frac{\varphi }{\sqrt{6}}}\right),
\end{equation}
Consider first the case for which the $F(R)$ gravity is given by
(\ref{caseonefrnontriv1}), so by combining Eqs. (\ref{r1}) and
(\ref{potentialvsigma}), the resulting Einstein frame potential is,
\begin{equation}\label{akyron1}
V(\varphi)=-\frac{1}{2} \Lambda  e^{-2 \sqrt{\frac{2}{3}} \varphi
}+4 \mu ^2-\frac{1}{8} 27 \mu ^2 n^4 e^{-2 \sqrt{\frac{2}{3}}
\varphi }-27 \mu ^2 n^3 e^{-\sqrt{\frac{3}{2}} \varphi }-36 \mu ^2
n^2 e^{-\sqrt{\frac{2}{3}} \varphi }-\mu ^2 n e^{-\frac{\varphi
}{\sqrt{6}}}\, ,
\end{equation}
which cannot be equal to the potential of Eq.
(\ref{smallalphaemodelpot}), regardless of the value of the
parameter $\Lambda$. The same applies if we choose the solution
(\ref{r2}), for the $F(R)$ gravity of Eq. (\ref{caseonefrnontriv1}).
So the only $F(R)$ gravity with physical interest which reproduces
correctly the Einstein frame potential (\ref{potentialvsigma}) is
that of Eq. (\ref{caseonefrnontriv}), which for the solution
(\ref{r2}) it becomes,
\begin{equation}\label{einsteinpotentunbroken}
V(\varphi )=-\frac{1}{2} \Lambda e^{-2 \sqrt{\frac{2}{3}} \varphi
}+4 \mu ^2+\frac{27}{8} \mu ^2 n^4 e^{-2 \sqrt{\frac{2}{3}} \varphi
}-8 \mu ^2 n e^{-\frac{\varphi }{\sqrt{6}}}\, ,
\end{equation}
so by choosing the constant $\Lambda$ to be equal to
$\Lambda=\frac{27 \mu ^2 n^4}{4}$, the potential
(\ref{einsteinpotentunbroken}) is identical to the potential of Eq.
(\ref{potentialvsigma}). Now let us find an approximate expression
for the Hubble rate during the slow-roll era, and in order to do
this, we use the expressions for the $F(R)$ gravity and its
derivative given in Eqs. (\ref{caseonefrnontriv}) and
(\ref{newfrnontrivs1}), and plug this in the first equation of Eq.
(\ref{cosmoeqns}), so after some algebra we obtain,
\begin{align}\label{simslowroll1}
& \frac{\Lambda}{H(t)^2}+\frac{27 H'(t)}{4 \mu ^2}+\frac{3 \sqrt{3}
H(t) \sqrt{\mu ^6 n^2}}{2 \mu ^4}+\frac{9 \sqrt{3} n^2 \sqrt{\mu ^6
n^2}}{2 \mu ^2 H(t)}+\frac{27 n^2}{4}\\ \notag & +\frac{9 \sqrt{3}
\sqrt{\mu ^6 n^2} H''(t)}{8 \mu ^4 H(t)^2}+\frac{9 H''(t)}{4 \mu ^2
H(t)}+\frac{3 \sqrt{3} \sqrt{\mu ^6 n^2} H'(t)}{\mu ^4 H(t)}+\frac{9
H'(t)^2}{8 \mu ^2 H(t)^2}=0\, ,
\end{align}
and therefore in the slow-roll limit, this differential equation
becomes,
\begin{equation}\label{tenfrlnewsec}
\frac{27 H'(t)}{4 \mu ^2}+\frac{3 \sqrt{3} H(t) \sqrt{\mu ^6 n^2}}{2
\mu ^4}+\frac{27 n^2}{4}=0\, .
\end{equation}
The above differential equation can be solved analytically, so the
resulting solution is,
\begin{equation}\label{mainalpha4solution}
H(t)\simeq \mathcal{C}e^{-Hi (t-t_k)}-H_0\, ,
\end{equation}
where the parameter $\mathcal{C}$ is an arbitrary integration
constant, $t_k$ is the horizon crossing time, while $H_i$ and $H_0$
are equal to,
\begin{equation}\label{ho}
H_0=\frac{3}{2} \sqrt{3} \mu  n,\,\,\,H_i=\frac{2 \mu  n }{3
\sqrt{3}}\, .
\end{equation}
Note that since the cosmic time in the solution
(\ref{mainalpha4solution}) takes values of the order $\sim
\mathcal{O}(10^{-20})$sec, the exponential is of the order $\sim
\mathcal{O}(1)$, so practically the evolution is a nearly de Sitter
evolution. Also, in order for the Hubble rate to to have negative
values, the parameter $\mathcal{C}$ must satisfy $\mathcal{C}\gg
H_0$. However, the evolution is actually a quasi de Sitter
expansion, and this can be seen by expanding the exponential in
(\ref{mainalpha4solution}) in powers of the cosmic time, so the
evolution becomes,
\begin{equation}\label{mainalpha4solution1}
H(t)\simeq \mathcal{C}-H_0-\mathcal{C}H_i (t-t_k)\, .
\end{equation}
Hence by having the evolution (\ref{mainalpha4solution1}) at hand,
we can easily calculate the observational indices. Following the
steps of the $\alpha=1/4$ case, the spectral index of the primordial
curvature perturbations in terms of the $e$-foldings number $N$, is
at leading order,
\begin{equation}\label{spectrscatotensornewsec}
n_s\simeq 1-\frac{4 C H_i}{\left(C \left(\frac{H_i
N}{C-H_0}-1\right)+H_0\right)^2},
\end{equation}
while the scalar-to-tensor ratio is,
\begin{equation}\label{spectrscatotensornewsec1}
r\simeq \frac{48 \mathcal{C}^2 H_i^2}{\left(-\frac{\mathcal{C} H_i
N}{\mathcal{C}-H_0}+\mathcal{C}-H_0\right)^4}.
\end{equation}
Therefore, in the large $N$ limit, the observational indices read,
\begin{equation}\label{spectrscatotensor1newsec}
n_s\simeq 1-\frac{2}{N},\, \,\,r=\frac{12}{N^2}\, ,
\end{equation}
where we have used the fact that during the slow-roll era, $N\simeq
\frac{(\mathcal{C}-H_0)^2}{2 \mathcal{C} H_i}$. So the resulting
observational indices are identical to the ones of Eq.
(\ref{jordanframeattract}), which means that the $R^2$ model in the
slow-roll approximation is the attractor of this $\alpha$ model too,
in the large $N$ limit of course.

As we did in the previous case, we need to examine whether the value
$\alpha=4$, yields viable results also in the Einstein frame. So we
examine the Einstein frame observational indices of Eq.
(\ref{observslowroli1}), for $\alpha=4$ in this case. For the set of
values $(N,\alpha)=(60,4)$, we obtain $n_s\simeq 0.966667$ and also
$r\simeq 0.0133333$, which are compatible with the Planck data of
Eq. (\ref{planckdata}). In addition, for the set of values
$(N,\alpha)=(50,1/4)$ we obtain $n_s\simeq 0.966667$, and the
predicted scalar-to-tensor ratio is $r\simeq 0.0192$, so these are
also compatible to the Planck constraints. Therefore, the case
$\alpha=4$ yields physically viable observational indices, for all
the physically relevant values of the $e$-foldings number $N$, which
lie in the interval $N=(50,60)$. However, the case $\alpha>4$ is
somewhat more involved and certain constraints should be imposed on
$\alpha$, in order for the Einstein frame observables to be
compatible with the observational data. For example if $N=60$, the
parameter $\alpha$ should satisfy $\alpha<29$, and for $N=50$, the
parameter $\alpha$ should satisfy $\alpha<20$. Nevertheless we will
not further discuss these cases, since it is difficult to obtain
analytical solutions in the Jordan frame, for these values of
$\alpha$.

\subsection{Limiting Cases of $\alpha$}

 However, let us briefly discuss the small-$\alpha$ limit of the algebraic equation (\ref{algegeneralalpha}),
 in order to see how the $F(R)$ gravity behaves. We shall present only the leading order behavior.
 Let us start with the leading order result, in which case, the algebraic equation (\ref{algegeneralalpha}) in the limit $\alpha \ll 1$ becomes,
\begin{equation}\label{limalgrbea}
R F_R^{\sqrt{\frac{1}{\alpha }}}-4 \alpha  \mu ^2
\left(F_R^{\sqrt{\frac{1}{\alpha }}}+\left(\sqrt{\frac{1}{\alpha
}}-2\right) n\right)=0\, ,
\end{equation}
so the solution to this equation is \cite{slowrollsergei},
\begin{equation}\label{kkouanalyticleading}
F_R(R)=\frac{R}{4\alpha\mu^2}+n(2-\frac{1}{\sqrt{2}})R^{1-\frac{1}{\sqrt{\alpha}}}\,
,
\end{equation}
which is valid for $0<\frac{1}{\sqrt{\alpha}}$. By integrating, we
find at leading order that the resulting $F(R)$ gravity is equal to,
\begin{equation}\label{kkouanalyticleadingsmallalpha}
F(R)=\frac{R^2}{8\alpha\mu^2}+n R^{2-\frac{1}{\sqrt{\alpha}}}\, .
\end{equation}
Since $R\gg \gamma$, the leading order that controls the dynamical
evolution in the small-$\alpha$ limit is the term $\sim R^2$,
therefore that is $R^22$ inflation what drives the evolution. It is
interesting that the inflationary $F(R)$ gravity of Eq.
(\ref{kkouanalyticleadingsmallalpha}) reminds the sector of unified
inflation-dark energy $F(R)$ gravity studied in Refs.
\cite{reviews1,eli1}. Then, by adding to above approximate
expression for alpha-attractor $F(R)$ inflationary theory the
exponential dark energy sector $F(R)=R-2\Lambda (1-e^{R/R_0})$, we
get unification of alpha-attractor inflation with dark energy in
$F(R)$ gravity.

We can easily find the observational indices for the $F(R)$ gravity
of Eq. (\ref{kkouanalyticleadingsmallalpha}), so by using the first
equation in Eq. (\ref{cosmoeqns}), upon differentiation with respect
to the cosmic time and by keeping leading order terms in the
slow-roll approximation we get the following differential equation,
\begin{equation}\label{oneofthefinals}
\frac{9 \gamma ^2 H(t) H^{(3)}(t)}{\mu ^2}+\frac{27 \gamma ^2 H(t)^2
H''(t)}{\mu ^2}+\frac{54 \gamma ^2 H(t) H'(t)^2}{\mu ^2}=0\, ,
\end{equation}
and by dividing with $H(t)^2$ we get,
\begin{equation}\label{enyaex}
\frac{9 \gamma ^2 H^{(3)}(t)}{\mu ^2 H(t)}+\frac{27 \gamma ^2
H''(t)}{\mu ^2}+\frac{54 \gamma ^2 H'(t)^2}{\mu ^2 H(t)}=0\, .
\end{equation}
So the dominant term in the slow-roll approximation is the second,
and by solving the resulting differential equation we obtain the
following Hubble rate,
\begin{equation}\label{hubbnewslw}
H(t)\simeq C_1-C_2 t\, ,
\end{equation}
which is valid during the slow-roll era. Hence the resulting
evolution is a quasi-de Sitter evolution, and by using the same line
of research as in the previous sections, the resulting observational
indices are identical to the ones appearing in Eqs.
(\ref{spectrscatotensor1newsec}) and (\ref{jordanframeattract}). So
actually, the $R^2$ model is the attractor of all the $F(R)$ gravity
equivalent theories of the Einstein frame $\alpha$-attractor models,
always in the slow-roll approximation. It is easy to see that due to
the relation (\ref{can}), the Ricci scalar as a function of the
canonical scalar field will be
$R=\frac{1}{A}e^{-\sqrt{\frac{2}{3\alpha}}\varphi}$, so a leading
order term in the potential is,
\begin{equation}\label{vpotleading}
V(\varphi)\sim
\frac{\sqrt{\alpha}}{A}e^{-\sqrt{\frac{2}{3\alpha}}\varphi}\, ,
\end{equation}
so indeed, the potential of Eq. (\ref{smallalphaemodelpot}) is
partially reconstructed. As it can be easily checked the leading
order term in the potential (\ref{vpotleading}) is generated by the
$R^2$ term in the $F(R)$ gravity.

Let us discuss another limiting case of $\alpha$, in which case
$\alpha$ is too large. In the large-$\alpha$ limit, the algebraic
equation (\ref{algegeneralalpha}) becomes approximately,
\begin{equation}\label{limalgrbealargealpha}
F_R R-4 \alpha  F_R^2 \mu ^2 (1-2 n)=0\, ,
\end{equation}
so the resulting $F(R) $ gravity is at leading order,
\begin{equation}\label{forevone}
F(R)\simeq \frac{R^2}{8 \mu ^2 (\alpha -2 \alpha  n)}\, .
\end{equation}
Therefore in this case too, the $R^2$ model is the attractor of the
$F(R)$ gravities, at least when the slow-roll approximation.

The resulting behavior in all the $F(R)$ cases we studied indicates
that the $F(R)$ gravity equivalent theories of the Einstein frame
$\alpha$-attractors models, have a unique attractor when the
slow-roll limit is used, and this is the $R^2$ model. Therefore,
regardless how the $F(R)$ gravity looks at second to leading order,
the observational indices are affected mainly by the leading order
term in the $F(R)$ gravity, and this is the reason why the resulting
observational indices are identical to the ones of the Starobinsky
model.

\section{Conclusions}

In this paper we investigated the $F(R)$ gravity equivalent theory
of some classes of Einstein frame $\alpha$-attractors models. The
full analytic treatment of the problem is not possible, so we chose
a convenient Einstein frame $\alpha$-attractor model, and we
calculated in detail the slow-roll indices in the slow-roll limit of
the $F(R)$ gravity theory. As we demonstrated, in the Jordan frame,
the attractors picture remains, since the resulting spectral index
of primordial curvature perturbations and the scalar-to-tensor ratio
remain attractors of the conveniently chosen $F(R)$ models.
Interestingly enough, the resulting observational indices in the
Jordan frame, are identical to the indices of the Starobinsky model,
and actually the $R^2$ model is the attractor in the Jordan frame,
at least when the slow-roll approximation is used. This result is
not accidental, since in all the cases we studied, the $F(R)$
gravity in the limit $R\gg 1$, are approximately equal to $\sim
R^2$, so the behavior is similar to the $R^2$ model.

An important issue is that the scalar-to-tensor ratio in the Jordan
frame does not depend on the ``$\alpha$'' parameter, and this is a
difference between the Einstein and Jordan frame description. The
question then is, does this occur due to the fact that the slow-roll
approximation is used? Do the slow-roll conditions in the two frames
impose different conditions on the resulting evolution? The quick
answer is no, the two frames are equivalent, but this can be shown
explicitly only for the $R^2$ model. However, for the simple
$\alpha$-model we used, we showed that this is not true, so the
question is if this holds true in general, or it holds true only for
the specific models we studied. We defer this task in the future
since the lack of analyticity forbids us for the moment to have a
definite answer on this.

\section*{Acknowledgments}

This work is supported by MINECO (Spain), project
 FIS2013-44881 (S.D.O) and by Min. of Education and Science of Russia (S.D.O
and V.K.O).

\section*{Appendix: The Case of an $\alpha$-dependent Canonical Transformation}

Consider the case that the canonical transformation which connects
the Einstein and the Jordan frame is $\alpha$-dependent. In this
case, the transformation (\ref{can}) becomes,
\begin{equation}\label{canalpha}
\varphi =\sqrt{\frac{3\alpha}{2}}\ln (F'(A))
\end{equation}
In this case, the metric in the Einstein and Jordan frames, namely
$\hat{g}_{\mu \nu}$ and $g_{\mu \nu}$, are related as follows,
\begin{equation}\label{conftransmetralpha}
g_{\mu \nu}=e^{-\sqrt{\frac{2}{3\alpha}}\varphi }\hat{g}_{\mu \nu
}\, ,
\end{equation}
where $g_{\mu \nu}$ denotes the Einstein frame metric. In order to
make the presentation more transparent, we will adopt another
notation different from the one we used in the main text. Suppose
that we identify $\psi^2=e^{\sqrt{\frac{2}{3\alpha}}\varphi }$, then
the Ricci scalar transforms as,
\begin{equation}\label{Ricciconftrans}
R=\psi^2\left( \tilde{R}+6\tilde{\square}\Psi-6\tilde{g}^{\mu
\nu}\partial_{\mu}\Psi\partial_{\nu}\Psi\right)\, ,
\end{equation}
where $\Psi=\ln \psi$. We can rewrite the Jordan frame $F(R)$ action
(\ref{pure}) as follows,
\begin{equation}\label{actionpuretranpro}
\mathcal{S}=\int \mathrm{d}^4x\sqrt{-g}\left(
\frac{1}{2}F'(R)R-V\right)\, ,
\end{equation}
where the potential $V$ is equal to,
\begin{equation}\label{potentialeqn}
V=\frac{F'R-F}{2}\, .
\end{equation}
The determinant of the metric under the transformation
(\ref{conftransmetralpha}) transforms as
$\sqrt{-g}=\psi^{-4}\sqrt{\tilde{g}}$, where in terms of $\psi$, the
scalar field is written as $\varphi=2\sqrt{\frac{3\alpha}{2}}\ln
\psi$. By combining the above, the resulting Einstein frame action
reads,
\begin{equation}\label{profinal}
\mathcal{S}=\int \mathrm{d}^4x \sqrt{-\tilde{g}}\left(
\tilde{R}-3\tilde{g}^{\mu \nu}\partial_{\mu}\Psi\partial_{\nu}\Psi-V
\right)\,.
\end{equation}
It can be easily shown that $\Psi=\varphi/\sqrt{6\alpha}$, so the
Einstein action can be written in terms of the scalar field
$\varphi$, and we have,
\begin{equation}\label{profinal1}
\mathcal{S}=\int \mathrm{d}^4x \sqrt{-\tilde{g}}\left(
\tilde{R}-\frac{1}{2\alpha}\tilde{g}^{\mu
\nu}\partial_{\mu}\varphi\partial_{\nu}\varphi-V(\varphi) \right)\,,
\end{equation}
where in this case the potential $V(\varphi)$ is equal to,
\begin{equation}\label{potentialfinalform}
V(\varphi)= \frac{1}{2}\left ( e^{-\sqrt{2/(3\alpha )}\varphi
}R\left (e^{\sqrt{2/(3\alpha )}\varphi} \right )-
e^{-2\sqrt{2/(3\alpha )}\varphi }F\left [ R\left
(e^{\sqrt{2/(3\alpha )}\varphi} \right ) \right ]\right )
\end{equation}
Hence by looking the resulting scalar action (\ref{profinal1}), it
can be seen that the $\alpha$-dependent conformal transformation
leads to a non-canonical scalar-tensor theory. Note that by further
re-scaling the scalar field $\varphi\to \phi\sqrt{\alpha}$, one
obtains a canonical scalar field theory, however in this case, the
canonical transformation (\ref{canalpha}) becomes,
 \begin{equation}\label{canalphanewsqu}
\phi =\sqrt{\frac{3}{2}}\ln (F'(A))\, ,
\end{equation}
which does not depend on $\alpha$ and therefore it is identical and
it leads to the same results as the transformation we used in Eq.
(\ref{can}).

\end{document}